\documentclass[12pt]{article}
\usepackage{epsfig,amsfonts,amssymb}
\usepackage{hyperref}
\usepackage{cite}
\input epsf.sty
\usepackage{cite}

\def\ZZZ{{\hbox{ Z\kern-1.6mm Z}}}
\def\RRR{{\hbox{ R\kern-2.4mm R}}}
\def\CCC{{\hbox{ C\kern-2.0mm C}}}
\def\zzz{{\hbox{z\kern-1mm z}}}

\newcommand{\qeq}{{\hbox{=\kern-2.3mm ? \kern.5mm }}}
\renewcommand{\qeq}{=}

\newcommand{\BBB}{{\cal B}}

\newcommand{\OO}{{\cal O}}

\newcommand{\NN}{{\cal N}}

\newcommand{\ba}{\bar a}
\newcommand{\bc}{\bar c}
\newcommand{\bd}{\bar d}
\newcommand{\bb}{\bar b}

\newcommand{\cp}{\check\Phi}
\newcommand{\crh}{\check\rho}
\newcommand{\cs}{\check\sigma}
\newcommand{\cv}{\check v}

\newcommand{\be}{\begin{equation}}
\newcommand{\ee}{\end{equation}}
\newcommand{\ben}{\begin{eqnarray}\displaystyle}
\newcommand{\een}{\end{eqnarray}}

\newcommand{\bea}[1]{\begin{eqnarray}\label{#1} }
\newcommand{\eea}{\end{eqnarray}}

\newcommand{\refb}[1]{(\ref{#1})}

\def\one{{\hbox{ 1\kern-.8mm l}}}
\def\zero{{\hbox{ 0\kern-1.5mm 0}}}

\begin{document}

\baselineskip 24pt

\begin{center}
{\Large \bf
Partition Functions of Torsion $>1$ Dyons in Heterotic
String Theory on $T^6$}

\end{center}

\vskip .6cm
\medskip

\vspace*{4.0ex}

\baselineskip=18pt

\centerline{\large \rm  Shamik Banerjee,  Ashoke Sen
and Yogesh K. Srivastava}

\vspace*{4.0ex}

\centerline{\large \it Harish-Chandra Research Institute}

\centerline{\large \it  Chhatnag Road, Jhusi,
Allahabad 211019, INDIA}

\vspace*{1.0ex}
\centerline{E-mail:  bshamik, sen, yogesh@mri.ernet.in}

\vspace*{5.0ex}

\centerline{\bf Abstract} \bigskip

The original proposal of Dijkgraaf, Verlinde and Verlinde for the
quarter BPS dyon partition function in heterotic string theory on
$T^6$ is known to correctly produce the degeneracy of dyons of
torsion 1, \i.e.\ dyons for which gcd($Q\wedge P$)=1. We propose
a generalization of this formula for dyons of arbitrary torsion. Our
proposal satisfies the constraints coming from S-duality invariance,
wall crossing formula, black hole entropy and the gauge theory
limit. Furthermore using our proposal we derive a 
general wall
crossing formula that is valid even 
when both the decay products are
non-primitive half-BPS dyons.

\vfill \eject

\baselineskip=18pt



\noindent{\bf Introduction:} Since the original proposal of
Dijkgraaf, Verlinde and Verlinde\cite{9607026}
for quarter BPS dyon spectrum
in heterotic string theory compactified on $T^6$, 
there has been
extensive study of dyon spectrum in a variety of 
$\NN=4$ supersymmetric string 
theories\cite{0412287,0505094,
0506249,0508174,0510147,0602254,
0603066,0605210,0607155,0609109,0612011,0702141,
0702150,0705.1433,0705.3874,0706.2363,
0708.1270} and also in $\NN=8$ and $\NN=2$ supersymmetric
string theories\cite{0506151,0711.1971}. 
However it has
been realised for some time that even in heterotic string theory
on $T^6$ the proposal of \cite{9607026} 
gives the correct dyon spectrum only
for a subset of dyons, -- those with unit torsion, \i.e.\
for which the electric and magnetic
charge vectors $Q$ and $P$ satisfy 
gcd($Q\wedge P$)=1\cite{0702150,0712.0043,0801.0149}.
In a previous paper we proposed a general set of constraints which
must be satisfied by the partition function of quarter BPS dyons in any
$\NN=4$ supersymmetric string theory and used these constraints
to propose a candidate for the dyon partition function for torsion
two dyons in heterotic string theory on 
$T^6$\cite{0802.0544}. In this
paper we extend our analysis to dyons of arbitrary torsion and propose
a form of the partition function of such dyons.

\noindent{\bf Proposal for the partition function:}
We consider the set $\BBB$ of all dyons of charge vectors $(Q,P)$ 
in heterotic string theory on $T^6$, with $Q$ being
$r$ times a primitive vector, $P$ a primitive vector and
$Q/r$ and $P$ admitting a primitive embedding in the Narain
lattice\cite{narain,nsw}, \i.e.\ all lattice vectors lying in the
plane of $Q$ and $P$ can be expressed as integer linear combinations
of $Q/r$ and $P$.
These dyons have torsion $r$, \i.e.\ gcd($Q\wedge P)=r$.
It was shown in \cite{0712.0043,0801.0149} 
that given any pair $(Q,P)$ of this type with
the same values of $Q^2$, $P^2$ and $Q\cdot P$, they are
related by T-duality transformation.
We denote by $d(Q,P)$ the index 
measuring the number of bosonic
supermultiplets minus the number of fermionic supermultiplets
of quarter BPS dyons carrying charges $(Q,P)$ -- up to a normalization
this can be identified with the helicity supertrace $B_6$ introduced in
\cite{9708062}. 
T-duality invariance of the theory tells us that $d(Q,P)$ must
be a function of the T-duality invariants, and hence has the form
$f(Q^2,P^2,Q\cdot P)$.
Then the dyon
partition function $1/\cp(\crh,\cs,\cv)$ is defined as
\be \label{e0}
{1\over \cp(\crh,\cs,\cv)} =\sum_{Q^2, P^2,Q\cdot P}
(-1)^{Q\cdot P+1} f(Q^2,P^2,Q\cdot P)
\, e^{i\pi (\cs Q^2 + \crh P^2 
+ 2 \cv Q\cdot P)}\, .
\ee
The sum in \refb{e0} runs over all possible values of $Q^2$, $P^2$
and $Q\cdot P$ in the set $\BBB$. This in particular requires
\be \label{e0.1}
Q^2/2\in r^2\ZZZ, \qquad P^2/2\in\ZZZ, \qquad Q\cdot P\in r\ZZZ
\, .
\ee
 The imaginary parts of 
 $(\crh,\cs,\cv)$ in \refb{e0} need to be adjusted to lie in a region
where the sum is
convergent. Although the index $f$ and hence the partition
function
$1/\cp$ so defined could depend on the
domain of the 
asymptotic moduli space
of the theory
in which we are computing the partition 
function\cite{0702141,0708.1270,0801.0149,0802.0544}, in
all known examples the dependence of $\cp$
on the domain
is found to come through the region of the complex
$(\crh,\cs,\cv)$ plane in which the sum is convergent. Thus 
\refb{e0} computed 
in different domains in the asymptotic moduli space
of the theory describes the same analytic
function $\cp$ in different domains in the complex 
$(\crh,\cs,\cv)$ plane.
We shall assume that the same feature holds for the partition function
under consideration. 

Since the quantization laws of $Q^2$, $P^2$
and $Q\cdot P$ imply that $\cp(\crh,\cs,\cv)$ is periodic under
independent shifts of $\crh$, $\cs$ and $\cv$ by 1, $1/r^2$ and 
$1/r$
respectively, eq.\refb{e0} can be inverted as
\ben \label{e0.2}
d(Q, P)
&=& {(-1)^{Q\cdot P+1}} r^3
\int_{i M_1-1/2}^{iM_1+1/2} d\crh
\int_{iM_2-1/(2r^2)}^{i M_2+1/(2r^2)} d\cs 
\int_{i M_3-1/(2r)}^{i M_3+1/(2r)} d\cv \, \nonumber \\
&& \qquad \qquad
e^{-i\pi (\cs Q^2 + \crh P^2 
+ 2 \cv Q\cdot P)} \, 
{1\over \cp(\crh, \cs, \cv)}\, ,
\een
provided the imaginary parts $M_1$, $M_2$ and $M_3$ of 
$\crh$, $\cs$ and $\cv$ are fixed in a region where the
original sum \refb{e0} is convergent.

Our proposal for $\cp(\crh,\cs,\cv)$ is
\be \label{e1}
\cp(\crh,\cs,\cv)^{-1}
= \sum_{\displaystyle{s\in\zzz, s|r}\atop 
\displaystyle{\bar s \equiv r/s}} g(s) \, 
{1\over \bar s^3} \, \sum_{k=0}^{\bar s^2-1} \, 
\sum_{l=0}^{\bar s-1}\, 
\Phi_{10}\left(\crh, s^2\cs +{k\over \bar s^2}, s\cv +{l\over \bar s}
\right)^{-1}\, ,
\ee
where 
\be \label{efs}
g(s)=s\, ,
\ee
and $\Phi_{10}(\crh,\cs,\cv)$ is the weight 10 Igusa cusp form
of $Sp(2,\ZZZ)$.
The sum over $k$ and $l$ in \refb{e1} makes $\cp$ periodic
under $\cs\to \cs +(1/r^2)$ and $\cv\to \cv + (1/r)$
as required. Even though the function $g(s)$ has a simple form
given in \refb{efs}, we shall carry out our analysis keeping
$g(s)$ arbitrary so that we can illustrate at the end how we fix the
form of $g(s)$ from the known wall crossing formula for decay into
a pair of primitive half-BPS dyons. In particular
we shall show that 
across a wall of marginal stability associated with the decay
of the original quarter BPS dyon
into a pair of half BPS states carrying charges
$(Q_1,P_1)$ and $(Q_2,P_2)$ with $(Q_1,P_1)$ being $N_1$
times a primitive lattice vector and $(Q_2,P_2)$ being $N_2$ times
a primitive lattice vector, the index jumps by an amount
\be \label{etot3}
\Delta d(Q,P)) = (-1)^{Q_1\cdot P_2-Q_2\cdot P_1+1}
 \, (Q_1\cdot P_2-Q_2\cdot P_1)\, 
\left\{\sum_{L_1|N_1} d_h\left({Q_1\over L_1}, 
{P_1\over L_1}\right) \right\}
\left\{\sum_{L_2|N_2} 
\, d_h\left({Q_2\over L_2}, {P_2\over L_2}\right)\right\}
\ee
for the choice $g(s)=s$ in \refb{e1}. 
Here $d_h(q,p)$ denotes the index of half BPS states carrying charges
$(q,p)$.
When $N_1=N_2=1$ both
the decay products are primitive and \refb{etot3} reduces to the
standard wall crossing formula\cite{0005049,0010222,0101135,0206072,
0304094,0702141,0702146,0706.3193,kont}.

Substituting \refb{e1} into \refb{e0.2}, extending the ranges
of $\cs$ and $\cv$ integral with the help of the sums over $k$
and $l$, and using the periodicity of $\Phi_{10}$ under
integer shifts of its arguments 
we can get a simpler expression for the index:
\ben \label{e1.1}
d(Q,P) &=&
{(-1)^{Q\cdot P+1}} \sum_{s|r} g(s)\, s^3\, 
\int_{i M_1-1/2}^{iM_1+1/2} d\crh
\int_{iM_2-1/(2s^2)}^{i M_2+1/(2s^2)} d\cs 
\int_{i M_3-1/(2s)}^{i M_3+1/(2s)} d\cv \, \nonumber \\ &&
e^{-i\pi (\cs Q^2 + \crh P^2 
+ 2 \cv Q\cdot P)} \, 
\Phi_{10}\left(\crh, s^2\cs 
, s\cv 
\right)^{-1}\, .
\een

The set of dyons considered above contains only a subset of dyons
of torsion $r$. This subset is known to be invariant under a 
$\Gamma^0(r)$ subgroup of the S-duality 
group\cite{0801.0149}. This requires
$\cp$ to be invariant under the transformation\cite{0802.0544}
\ben \label{es.1}
\cp(\crh',\cs',\cv') = \cp(\crh,\cs,\cv) \quad &\hbox{for}& \quad
\pmatrix{\crh' &\cv'\cr \cv' & \cs'} 
= \pmatrix{d & b\cr c & a} \pmatrix{\crh & \cv\cr \cv &\cs}
\pmatrix{d & c\cr b & a}\, , \nonumber \\ \cr
&&  a,c,d\in\ZZZ, \quad
b\in r\ZZZ, \quad ad-bc=1\, .
\een
On the other hand a general S-duality transformation matrix
$\pmatrix{a & b\cr c & d}$ outside $\Gamma^0(r)$ will take us
to dyons of torsion $r$ outside the set $\BBB$\cite{0801.0149}.
Thus with the help of these S-duality transformations on
$\cp$ we can
determine the partition function for other torsion $r$ dyons lying
outside the set $\BBB$ considered above. In particular if we
consider the set of dyons carrying charges $(Q',P')$ related
to $(Q,P)$ via an S-duality transformation
\be \label{ess.1}
(Q,P)=(aQ'+bP', cQ'+dP'), \quad \pmatrix{a & b\cr c & d}\in
SL(2,\ZZZ)\, ,
\ee
and denote by $1/\cp'$ the partition function of these dyons, then
$\cp'$ is related to $\cp$ via the relation
\be \label{ess.2}
\cp'(\crh',\cs',\cv') = \cp(\crh,\cs,\cv) \quad \hbox{for} \quad
\pmatrix{\crh' &\cv'\cr \cv' & \cs'} 
= \pmatrix{d & b\cr c & a} \pmatrix{\crh & \cv\cr \cv &\cs}
\pmatrix{d & c\cr b & a} \, .
\ee
This allows us to determine the partition function of all other sets
of torsion $r$ dyons from the partition function given in
\refb{e1}. In particular for $r=2$, choosing $\pmatrix{a & b\cr c & d}
=\pmatrix{1 & 1\cr 0 & 1}$ we recover the dyon
partition function proposed  in \cite{0802.0544}.

We shall now show that the proposed partition function 
\refb{e1} satisfies
various consistency tests described in \cite{0802.0544}.

\noindent{\bf S-duality invariance:}
We shall first verify the required
S-duality invariance of the partition function described in
eq.\refb{es.1}.
Using $Sp(2,\ZZZ)$ invariance of $\Phi_{10}(x,y,z)$, and that
$b$ is a multiple of $r$ for $\pmatrix{a&b\cr c&d}\in
\Gamma^0(r)$ one can
show that
\be \label{es.2}
\Phi_{10}\left(\crh', s^2 \cs' +{k\over \bar s^2}, s\cv'
+{l\over \bar s}\right)
= \Phi_{10}\left(\crh, s^2 \cs +{k'\over \bar s^2}, s\cv
+ {l'\over \bar s}\right)\, ,
\ee
where
\be \label{es.3}
k' = k d^2 -2cdlr\in\ZZZ, \qquad l'=(ad+bc)l -bdk/r\in\ZZZ\, .
\ee
Thus
\be \label{es.4}
\sum_{k=0}^{\bar s^2-1} \, 
\sum_{l=0}^{\bar s-1}\, 
\Phi_{10}\left(\crh', s^2\cs' +{k\over \bar s^2}, s\cv' +{l\over \bar s}
\right)^{-1}
= \sum_{k'=0}^{\bar s^2-1} \, 
\sum_{l'=0}^{\bar s-1}\, 
\Phi_{10}\left(\crh, s^2 \cs +{k'\over \bar s^2}, s\cv
+ {l'\over \bar s}\right)^{-1}\, ,
\ee
and we have
the required relation \refb{es.1}.

\noindent{\bf Wall crossing formula:}
We shall now verify that \refb{e1.1} is consistent with the
wall crossing formula. As in
\cite{0802.0544} we shall only
consider the decay into a pair of half-BPS 
dyons\cite{0702141,0707.1563,0707.3035,0710.4533}
\be \label{e2}
(Q,P) \to (Q_1,P_1)+(Q_2,P_2)\, ,
\ee
\be \label{e2.55}
(Q_1,P_1) = (\alpha Q +\beta P, \gamma Q +\delta P), 
\qquad (Q_2,P_2)=
(\delta Q -\beta P, -\gamma Q + \alpha P)\, ,
\ee
\be \label{e3}
\alpha\delta=\beta\gamma, \qquad \alpha+\delta=1\, .
\ee
Since any lattice vector lying in the plane of $Q$ and $P$ 
can be expressed as a linear combination of
$Q/r$ and $P$ with integer coefficients, 
we must have $\beta,\delta,\alpha\in\ZZZ$, 
$\gamma\in
\ZZZ/r$. Thus we can
write $\gamma=\gamma'/K$, where $K\in\ZZZ$,
$K|r$ and gcd($\gamma',K$)=1.
The condition $\alpha\delta=\beta\gamma$ 
together with the integrality
of $\alpha,\beta,\delta$ now tells us that $\beta$ must be of the form
$K\beta'$ with $\beta'\in\ZZZ$. Thus we have
\be \label{e5}
\beta=K\, \beta', \quad
\gamma={\gamma'\over K}, \quad K,\alpha, \delta,
\beta',\gamma'\in\ZZZ, 
\quad K|r, \quad \gcd(\gamma',K)=1\, .
\ee
Using eqs.\refb{e3}, \refb{e5} we have
\be \label{est1}
\alpha+\delta =1, \qquad \alpha\delta =\beta'\gamma', \qquad
\alpha,\beta',\gamma',\delta \in\ZZZ\, .
\ee
The analysis of \cite{0702141} now shows that we can
find $a'$, $b'$, $c'$, $d'$ such that
\be \label{est2}
\alpha=a'd', \quad \beta'=-a'b', \quad \gamma'=c'd', \quad
\delta =-b'c', \quad a',b',c',d'\in\ZZZ, \quad a'd'-b'c'=1\, .
\ee
As a consequence of \refb{est2} and the relation
$\gcd(\gamma',K)=1$ we have
\be \label{egcd}
\gcd(a',b')=\gcd(a',c')=\gcd(c',d')=\gcd(b',d')=1, \quad
\gcd(c',K)=\gcd(d',K)=1\, .
\ee
Using eqs.\refb{e5}-\refb{est2} we can now express \refb{e2.55} as
\be \label{eq.1}
(Q_1,P_1) = (a'K,c') (d'\bar K Q/r - b'P), \qquad
(Q_2,P_2) = (b'K, d') (-c'\bar K Q/r+a'P), \qquad 
\bar K \equiv r/K\, .
\ee
Since according to \refb{egcd}, gcd($a'K,c'$)=1, gcd($b'K, d')$=1,
and since any lattice vector lying in the $Q$-$P$ plane can be
expressed as integer linear combinations of $Q/r$ and $P$, 
it follows from
\refb{eq.1} that $(Q_1,P_1)$ can be regarded as $N_1$ times
a primitive vector and $(Q_2,P_2)$ can be regarded as $N_2$ times
a primitive vector where
\be \label{eq.2}
N_1=\gcd(d'\bar K, b') = \gcd(\bar K, b'), \qquad
N_2=\gcd(c'\bar K, a') = \gcd(\bar K, a')\, .
\ee
In the last steps we have again made use of \refb{egcd}. It follows
from \refb{egcd} and \refb{eq.2} that
\be \label{e7}
\gcd(N_1,N_2)=1, \quad N_1N_2 | \bar K\,  .
\ee

We shall now use the formula \refb{e1.1} for the index in
different regions of the moduli space and calculate the change in the
index as we cross the wall of marginal stability associated with
the decay \refb{e2}. For this we need to know how to choose the
imaginary parts of $\crh$, $\cs$ and $\cv$ along the integration contour
for the two domains lying on the two sides of this wall of marginal
stability. A prescription for choosing this
 contour was postulated in
\cite{0802.0544} according 
to which as we cross the wall of marginal
stability associated with the decay \refb{e2}, the integration 
contour crosses a pole of the partition function at
\be \label{e4}
\crh \gamma - \cs \beta + \cv (\alpha-\delta) = 0\, .
\ee
Thus the change in the index can be calculated by evaluating the
residue of the partition function at this pole.
We shall now examine for which values of $s$ the 
$\Phi_{10}(\crh,
s^2\cs,s\cv)^{-1}$ term in the expansion \refb{e1.1} has a pole
at \refb{e4}. The poles of $\Phi_{10}(\crh,
s^2\cs,s\cv)^{-1}$ are known to be at
\ben \label{est3}
&& n_2\,  s^2\, (\crh\cs - \cv^2) 
+ n_1 s^2\cs -m_1\crh+m_2+ j 
s\cv =0
\nonumber \\
&& m_1,
n_1, m_2, n_2 \in \ZZZ, \quad j\in 2\ZZZ+1, \quad
m_1 n_1 + m_2 n_2 +\frac{j^2}{4} = {1\over 4}\, .
\een
Comparing \refb{est3} and \refb{e4} we see that we must have
\be \label{est4}
m_2=n_2=0, \qquad j ={\lambda\over s}(\alpha-\delta), \qquad
n_1=-{\lambda\over s^2}\beta, \qquad m_1=-\lambda\gamma\, ,
\ee
for some $\lambda$.
The last condition in \refb{est3}, together with eqs.\refb{e3} now
gives
\be \label{est5}
\lambda=s\, .
\ee
Thus we have from \refb{e5}, \refb{est4}
\be \label{est6}
j=\alpha-\delta, \qquad m_1=-s \gamma = -\gamma' s/K, \qquad
n_1=-\beta/s = -K\beta'/s\, .
\ee
Since $\gcd(\gamma', K)=1$, the second equation in \refb{est6}
shows that $s$ must be a multiple of $K$:
\be \label{est7}
s = K\, \tilde s, \qquad \tilde s\in\ZZZ\, .
\ee
Substituting this into the last equation in \refb{est6} and using
\refb{est2} we see that
\be \label{est8}
n_1={a'b'\over \tilde s} \quad \Rightarrow
\quad {a'b'\over \tilde s} \in \ZZZ\, .
\ee
Since gcd($a',b'$)=1, we must
have a unique decomposition
\be \label{est8.5}
\tilde s=L_1 L_2, \qquad L_1|b', \qquad L_2|a'\, .
\ee
On the other hand since $s$ divides $r$, it follows from
\refb{est7} that $\tilde s$ must divide
$r/K=\bar K$. Thus
\be \label{est9}
L_1|\bar K, \qquad L_2|\bar K\, .
\ee
It now follows from \refb{eq.2} that
\be \label{est10}
L_1|N_1, \qquad L_2|N_2\, .
\ee
Conversely, given any pair $(L_1,L_2)$ satisfying \refb{est10},
it follows from \refb{eq.2} that $L_1$, $L_2$ will satisfy
\refb{est8.5}, \refb{est9}. This allows us to find integers
$m_1$, $n_1$, $j$ satisfying \refb{est4} via 
eqs.\refb{est5}-\refb{est8.5}.

This shows that the poles
of $\cp(\crh,\cs,\cv)^{-1}$ at \refb{e4} can come from the 
$s=KL_1L_2$ terms in \refb{e1} for $L_1|N_1$ and
$L_2|N_2$.
Our next task is to find the 
residues at these poles to compute the change in the index as we
cross this wall.
For this we define
\be \label{ebardef}
\ba = a'/L_2, \quad \bd =d'L_2, \quad \bb=b'/L_1,
\quad \bc = c'L_1, \quad s_0=KL_1L_2\, .
\ee
It follows from \refb{est8.5} that $\ba,\bb,\bc,\bd$ are all
integers.
In terms of these variables
the location of the pole given in \refb{e4} can be expressed
as
\be \label{extrapole}
s_0^{-1}\,
\bc\bd \crh + \ba\bb s_0\cs +(\ba\bd +\bb\bc) \cv=0\, .
\ee
We now define
\be \label{ee.1}
\crh'=\bd^2 \crh + \bb^2 s_0^2 \cs + 2\bb\bd s_0 \cv, \quad
\cs' = \bc^2 \crh + \ba^2 s_0^2 \cs + 2\ba\bc s_0\cv, \quad
\cv' = \bc\bd \crh + \ba\bb s_0^2\cs +(\ba\bd +\bb\bc) s_0\cv\, .
\ee
The change of variables from $(\crh, s_0^2\cs, s_0\cv)$ to
$(\crh',\cs',\cv')$ is an $Sp(2,\ZZZ)$ transformation. Thus
we have
\be \label{ee.2}
\Phi_{10}(\crh,s_0^2\cs,s_0\cv)=\Phi_{10}(\crh',\cs',\cv')\, .
\ee
In the primed variables the desired pole at \refb{extrapole}
is at $\cv'=0$.
We also have
\be \label{ee.4}
\crh P^2 + \cs Q^2 + 2\cv Q\cdot P
= \crh' P^{\prime 2} + \cs' Q^{\prime 2}
+2\cv' Q'\cdot P'\, ,
\ee
where
\be \label{ee.5}
Q'= \bd \, Q/s_0 - \bb P, \qquad P'=-\bc Q/s_0 + \ba P\, .
\ee
Finally we have
\be \label{ee.3}
d\crh' d\cs' d\cv' = s_0^3 d\crh d\cv d\cs\, .
\ee
This is consistent with the fact that $\Phi_{10}(\crh',\cs',\cv')$
is invariant under integer shifts in $\crh'$, $\cs'$ and $\cv'$
so that in the primed variables the
volume of the unit cell is 1, while in the unprimed variables the
volume of the unit cell is $1/s_0^3$. We can now express the
change in the index from the
$s=s_0$ term in \refb{e1.1} as
\ben \label{ee.6}
(\Delta d(Q,P))_{s_0}
&=& 
{(-1)^{Q\cdot P+1}} \, g(s_0) \,  
\int_{i M_1'-1/2}^{iM_1'+1/2} d\crh'
\int_{iM_2'-1/2}^{i M_2'+1/2} d\cs' 
\ointop d\cv' \, \nonumber \\ &&
e^{-i\pi (\cs' Q^{\prime 2 }+ \crh' P^{\prime 2} 
+ 2 \cv' Q'\cdot P')} \, 
\Phi_{10}\left(\crh',  \cs'
, \cv' 
\right)^{-1}\, ,
\een
where the $\cv'$ contour is around the origin, -- 
as in \cite{0802.0544}
we shall use the convention that the contour is in the clockwise
direction. Using the fact that
\be \label{ee.7}
\Phi_{10}\left(\crh',  \cs'
, \cv' 
\right) = -4\pi^2 \, \cv^{\prime 2} \, \eta(\crh')^{24}\, 
\eta(\cs')^{24}
+ \OO\left(\cv^{\prime 4}\right)\, ,
\ee
near $\cv'=0$, we get
\ben \label{ee.8}
(\Delta d(Q,P))_{s_0}
&=& 
{(-1)^{Q\cdot P+1}} \, g(s_0) \,  Q'\cdot P'\, 
\int_{i M_1'-1/2}^{iM_1'+1/2} d\crh' 
e^{-i\pi  \crh' P^{\prime 2}}\, \eta(\crh')^{-24}\nonumber \\
&& \qquad \qquad \qquad 
\int_{iM_2'-1/2}^{i M_2'+1/2} d\cs' 
e^{-i\pi \cs' Q^{\prime 2 }}\, \eta(\cs')^{-24}\nonumber \\
&=& {(-1)^{Q\cdot P+1}} \, g(s_0) \,  Q'\cdot P'\, 
d_h(Q',0) \, d_h(P',0)\, , 
\een
where $d_h(q,0)$ denotes the index measuring the number
of bosonic half BPS supermultiplets minus the number of
fermionic half BPS supermultiplets carrying charge $(q,0)$.

We shall now express the right hand side of \refb{ee.8}
in terms of the charges $(Q_1,P_1)$ and $(Q_2,P_2)$
of the decay products. 
First of all it is easy to see that
\be \label{ek.1}
(-1)^{Q\cdot P} = (-1)^{Q_1\cdot P_2 - Q_2\cdot P_1}\, .
\ee
Furthermore it follows from \refb{eq.1}, \refb{ebardef}
and \refb{ee.5}
that
\be \label{ef.1}
(Q_1,P_1) = L_1\, \left(a'KQ',c'Q'\right), 
\qquad (Q_2,P_2)
 =L_2\, \left(b'KP',d'P'\right)
\, ,
\ee
\be \label{ee.9}
Q_1\cdot P_2 - Q_2\cdot P_1= s_0\, Q'\cdot P'  \, .
\ee
Now
according to \refb{egcd} the pair of integers
$\left(a'K,c'\right)$ are relatively prime, and
the pair of integers
$\left(b'K,d'\right)$ are also relatively prime.
Thus using S-duality invariance we can write
\ben \label{eh2}
d_h\left({Q_1\over L_1},{P_1\over L_1}\right)
&=& d_h\left(a'KQ', c'Q'\right)
 = d_h(Q',0), \nonumber \\
d_h\left({Q_2\over L_2},{P_2\over L_2}\right)
&=& d_h\left(b'KP' , d'P' \right)
= d_h(P',0)\, .
\een
Using \refb{ek.1}, \refb{ee.9} and \refb{eh2} 
we can now express
\refb{ee.8} as 
\be \label{eh3}
(\Delta d(Q,P))_{s_0}
=
{(-1)^{Q_1\cdot P_2-Q_2\cdot P_1+1}} 
\, g(s_0) \,  {1\over s_0} \, (Q_1\cdot P_2-Q_2\cdot P_1)\, 
d_h\left({Q_1\over L_1}, {P_1\over L_1}\right) 
\, d_h\left({Q_2\over L_2}, {P_2\over L_2}\right)\, .
\ee
For a given decay $K$ is fixed but $L_1$ and $L_2$ can vary over
all the factors of $N_1$ and $N_2$. Thus the total change in the
index is obtained by summing over all possible values of
$s_0$ of the form $KL_1L_2$. Thus gives
\ben \label{etot}
\Delta d(Q,P)) &=& (-1)^{Q_1\cdot P_2-Q_2\cdot P_1+1}
 \, (Q_1\cdot P_2-Q_2\cdot P_1)\, \nonumber \\ &&
\sum_{L_1|N_1} \sum_{L_2|N_2} \, g(KL_1L_2) \,
{1\over KL_1L_2}\, 
d_h\left({Q_1\over L_1}, {P_1\over L_1}\right) 
\, d_h\left({Q_2\over L_2}, {P_2\over L_2}\right)\, .
\een
We can now fix the form of the function $g(s)$ by considering a
decay where the decay products are primitive, \i.e.\ $N_1=N_2=1$.
In this case we have $L_1=L_2=1$, and \refb{etot} takes the
form\footnote{In this argument we have implicitly assumed that
for a given $K$, it is possible to find integers $a'$, $b'$, $c'$, $d'$
satisfying $a'd'-b'c'=1$, $\gcd(c',K)=\gcd(d',K)=\gcd(a',\bar K)
=\gcd(b',\bar K)=1$, so that \refb{egcd} holds and
we have $N_1=N_2=1$ according to
\refb{eq.2}. If either $K$ or $\bar K$ is 
odd then this assumption holds
with the choice $\pmatrix{a'&b'\cr c'&d'} =\pmatrix{2 & 1\cr 1&1}$
or $\pmatrix{1 & 1\cr 1 &2}$. If $K$ and $\bar K$
are both even then we cannot find $a'$, $b'$, $c'$ and $d'$
satisfying all the requirements since in order to satisfy
$a'd'-b'c'=1$ at least one of them must be even. However in this
case if we choose $\pmatrix{a'&b'\cr c'&d'}=\pmatrix{2 & 1\cr
1 & 1}$ then we satisfy \refb{egcd}
and have $N_1=1$ and $N_2=2$ according to
\refb{eq.2}. We can now demand
that the wall crossing formula in this case agrees with the one
derived in \cite{0802.0544} for decays where one of the
decay products is twice a primitive vector. This gives
$g(K)=K$ even for $K$, $\bar K$ both even.}
\be \label{et2}
\Delta d(Q,P))  = (-1)^{Q_1\cdot P_2-Q_2\cdot P_1+1}
 \, (Q_1\cdot P_2-Q_2\cdot P_1) \, g(K)\, {1\over K}\, 
 d_h(Q_1,P_1)\, d_h(Q_2,P_2)\, .
 \ee
 In order that this agrees with the standard 
 wall crossing formula
 for primitive 
 decay\cite{0005049,0010222,0101135,0206072,
 0304094,0702141,0702146,0706.3193}
 we must have $g(K)=K$. Since this result
 should hold for all $K|r$ we see that we must set $g(s)=s$ as given in
 \refb{efs}. Using this we can simplify the wall crossing 
 formula \refb{etot}
for generic non-primitive decays to the form given in
 \refb{etot3}.
 
\noindent{\bf Black hole entropy:} In order 
to reproduce the leading contribution to the black
hole entropy in the limit of large charges,   
the partition function must have
a pole  at\cite{9607026}
\be \label{ebh1}
\crh\cs -\cv^2 +\cv=0\, .
\ee
Furthermore, in order to reproduce
the black hole
entropy to first non-leading order, the inverse
of the partition function 
near this pole must
behave as\cite{0412287,0802.0544}
\be \label{ebh2}
\cp(\crh,\cs,\cv) 
 \propto (2v-\rho-\sigma)^{10}\,
\{ v^2 \, \eta(\rho)^{24}\, \eta(\sigma)^{24} 
+  \OO(v^4)\}\, ,
\ee
where
\be \label{ebh3}
\rho = {\crh \cs - \cv^2\over \cs}, \quad
\sigma = {\crh \cs - (\cv-1)^2\over \cs}, \quad
v= {\crh \cs - \cv^2+\cv\over \cs}\, .
\ee
We shall now examine the poles of $\cp$ given in \refb{e1} to
see if it satisfies the above relations. For this we recall 
eq.\refb{est3} giving the pole of 
$\Phi_{10}(\crh,s^2\cs,s\cv)^{-1}$. Comparing \refb{est3}
with \refb{ebh1} we see that in order to get a pole at
\refb{ebh1} we must choose in \refb{est3}
\be \label{ebh4}
n_2=\lambda/s^2, \qquad j=\lambda/s, \qquad n_1=m_1=m_2=0\, ,
\ee
for some $\lambda$. The requirement $m_1n_1+m_2n_2+{1\over 4}
j^2={1\over 4}$ then gives
\be \label{ebh5}
\lambda=s, \qquad n_2={1\over s}\, .
\ee
Since $n_2$ must be an integer this gives $s=1$. Thus only the
$s=1$ term in \refb{e1} 
has a pole at \refb{ebh1}. It follows from the 
known behaviour
of $\Phi_{10}$ near its zeroes that $\cp$ defined in \refb{e1}
satisfies the requirement \refb{ebh2}.

\noindent{\bf Gauge theory limit:} Finally we shall show that in
special regions of the Narain moduli space where the low lying
states in string theory describe a non-abelian gauge theory, the
proposed dyon spectrum of string theory reproduces the known
results in gauge theory. Since the T-duality invariant 
metric $L$ in the Narain moduli space descends to the
negative of
the Cartan metric in gauge theories, and since the Cartan metric
is positive definite, all the gauge theory dyons have the property
\be \label{egg1}
Q^2<0, \qquad P^2<0, \qquad Q^2P^2 > (Q\cdot P)^2\, .
\ee
Thus in order to identify dyons which could be interpreted as
gauge theory dyons in an appropriate limit we must focus on
charge vectors satisfying \refb{egg1}.

In order to identify such dyons we need to expand the partition
function $\cp^{-1}$ in powers of $e^{2\pi i\crh}$, $e^{2\pi i\cs}$
and $e^{2\pi i\cv}$ and pick up the appropriate terms in the
expansion. For this we need to identify a region in the $(\crh,\cs,\cv)$
space (or equivalently in the asymptotic moduli space) where we
carry out the expansion, since in different regions we have different
expansion. We shall consider the region
\be \label{egg2}
\Im(\crh),\Im(\cs) >> -\Im(\cv) >> 1\, ,
\ee
where $\Im(z)$ denotes the imaginary part of $z$. 
The results in other relevant regions 
will be related to the ones in this
region by S-duality transformation.
In the region \refb{egg2}
the only term in $\Phi_{10}\left(\crh,s^2\cs+{k\over \bar 
s^2},s\cv+{l\over \bar s}\right)^{-1}$ which has
powers of $e^{2\pi i\crh}$, $e^{2\pi i\cs}$
and $e^{2\pi i\cv}$ compatible with the requirement
\refb{egg1} is
\be \label{egg3}
e^{-2\pi i\crh - 2\pi i (s^2 \cs+{k\over \bar s^2}) 
-2\pi i (s\cv+{l\over \bar s})}\, .
\ee
This is in fact the leading term in the expansion in the
limit \refb{egg2}.
Substituting this into \refb{e1} and performing the sum over $k$
and $l$ we see that only the $s=r$ term in the sum survives and
the result is
\be \label{egg4}
 r\, e^{-2\pi i (\crh + r^2\cs+r\cv)}\, .
\ee
This corresponds to dyons with 
\be \label{egg4.5}
Q^2/2=-r^2, \quad P^2/2=-1, \quad
Q\cdot P=-r\, ,
\ee
with an index of $(-1)^{r+1} \, r$.

We can also determine the walls of marginal stability which border
the domain in which these dyons exist. This requires determining
the region in the $(\Im(\crh),\Im(\cs),\Im(\cv))$ space in which
the expansion  \refb{egg4} of $\Phi_{10}(\crh,r^2\cs,r\cv)^{-1}$
is valid since then we can determine the associated walls of marginal
stability using \refb{e4}. For this we shall utilize the
known results for $r=1$; in this case the walls of marginal stability
bordering the domain in which \refb{egg4} is valid correspond to
the decays into $(Q,0)+(0,P)$, $(Q-P, 0)+(P,P)$ and
$(0,P-Q)+(Q,Q)$ respectively\cite{0708.3715}. Using
\refb{e4} we now see that validity of the expansion of 
$\Phi_{10}(\crh,\cs,\cv)^{-1}$ given by \refb{egg4}
with $r=1$ is bounded by the following surfaces
in the $(\Im(\crh),\Im(\cs),\Im(\cv))$ space:
\be \label{egg5}
\Im(\cv)=0\, , \qquad \Im(\cv+\cs)=0, \qquad 
\Im(\cv+\crh)=0\, .
\ee
We can now simply scale $\cs$ by $r^2$ and $\cv$ by $r$ to
determine the region of validity of the expansion \refb{egg4}
for $\Phi_{10}(\crh,r^2\cs,r\cv)$:
\be \label{egg6}
\Im(\cv)=0\, , \qquad \Im(\cv+r\cs)=0, \qquad 
\Im(r\cv+\crh)=0\, .
\ee
Comparing these with \refb{e4} we now see that the corresponding
walls of marginal stability are associated with the decays into
\be \label{egg7}
(Q,0)+(0,P), \qquad (Q-rP, 0)+(rP,P), \qquad 
\left(0, P-{1\over r}Q\right)+
\left(Q,{1\over r}Q\right)\, .
\ee 

Let us now compare these results with dyons in $\NN=4$
supersymmetric $SU(3)$ gauge
theory. If we denote by $\alpha_1$ and $\alpha_2$ a pair of
simple roots of $SU(3)$ with $\alpha_1^2=\alpha_2^2=-2$ and
$\alpha_1\cdot\alpha_2=1$, then the analysis of 
\cite{9804174,9907090,0005275,0609055,0802.0761}
shows that the gauge theory contains dyons of charge
\be \label{egg8}
(Q,P) = (r\alpha_1, -\alpha_2)\, ,
\ee
with index $(-1)^{r+1}\, r$. These are precisely the dyons of the
type given in \refb{egg4.5}. Furthermore using string junction
picture\cite{9712211,9804160}, ref.\cite{9804174} also
determined the walls of marginal stability bordering the
domain in which these dyons exist. These also coincide with
\refb{egg7}.

The spectrum in gauge theory contains other dyons of torsion $r$
related to the ones described above by S-duality transformation. Since
our construction is manifestly S-duality invariant, the results for these
dyons can also be reproduced from the general formula
given in \refb{e1}.

Gauge theory also contains other dyons
which are not related to the ones described here by 
S-duality\cite{0005275,0609055}. These
typically require higher gauge groups and has additional fermionic
zero modes besides the ones required by broken supersymmetry.
Quantization of these additional zero modes gives 
rise to additional
bose-fermi degeneracy, and as a result the index being computed
here
vanishes for these dyons. This is also apparent from the fact
that these dyons typically exist only in a subspace of the full moduli
space; as a result when we move away from this subspace the
various states combine and become non-BPS. Some aspects of these
dyons have been discussed recently in \cite{0712.3625,0802.0761}.

\medskip

\noindent {\bf Acknowledgment:} We would like to thank
Nabamita Banerjee, Miranda Cheng, Atish Dabholkar,
Justin David, 
Frederik Denef, Dileep Jatkar, Suresh Nampuri,
K. Narayan and Sunil Mukhi for useful discussions.


\end{document}